\documentclass[%
 reprint,
 amsmath,amssymb,superscriptaddress,aps,
]{revtex4-2}

\usepackage{graphicx}
\usepackage{dcolumn}
\usepackage{bm}

\begin{document}

\preprint{APS/123-QED}

\title{Toward Monoatomic-Layer Field Confinement Limit via Acoustic Graphene Plasmons}

\author{In-Ho Lee}
\affiliation{Department of Electrical and Computer Engineering, University of Minnesota, Minneapolis, Minnesota 55455, USA}
\author{Tony Low}
\affiliation{Department of Electrical and Computer Engineering, University of Minnesota, Minneapolis, Minnesota 55455, USA}
\author{Luis Mart\'{i}n-Moreno}
\affiliation{Instituto de Nanociencia y Materiales de Aragón (INMA), CSIC-Universidad de Zaragoza, Zaragoza 50009, Spain}
\affiliation{Departamento de Física de la Materia Condensada, Universidad de Zaragoza, Zaragoza 50009, Spain}
\author{Phaedon Avouris}
\affiliation{Department of Electrical and Computer Engineering, University of Minnesota, Minneapolis, Minnesota 55455, USA}
\affiliation{IBM Thomas J. Watson Research Center, Yorktown Heights, New York 10598, USA}
\author{Sang-Hyun Oh}
\email{sang@umn.edu}
\affiliation{Department of Electrical and Computer Engineering, University of Minnesota, Minneapolis, Minnesota 55455, USA}

\date{December 25, 2020}

\begin{abstract}
Vertical plasmonic coupling in double-layer graphene leads to two hybridized plasmonic modes: optical and acoustic plasmons with symmetric and anti-symmetric charge distributions across the interlayer gap, respectively. However, in most experiments based on far-field excitation, only the optical plasmons are dominantly excited in the double-layer graphene systems. Here, we propose strategies to selectively and efficiently excite acoustic plasmons with a single or multiple nano-emitters. The analytical model developed here elucidates the role of the position and arrangement of the emitters on the symmetry of the resulting graphene plasmons. We present an optimal device structure to enable experimental observation of acoustic plasmons in double-layer graphene toward the ultimate level of plasmonic confinement defined by a monoatomic spacer, which is inaccesible with a graphene-on-a-mirror architecture. 

\end{abstract}

\maketitle


\section{Introduction}
Various two-dimensional (2D) materials including graphene\cite{zhang2005experimental,neto2009electronic,hwang2007dielectric, jablan2009plasmonics, koppens2011graphene, chen2012optical,fei2012gate,nikitin2011fields} and black phosphorus\cite{low2014plasmons, liu2016localized, lee2018anisotropic} have emerged as promising plasmonic platforms due to their ability to confine light into deep sub-diffractional volumes and modulate plasmon properties through doping. In particular, subwavelength light confinement via polaritons\cite{basov2016polaritons,low2017polaritons,basov2020polariton} in 2D materials has been a focus of intense research\cite{brar2013highly, woessner2015highly,caldwell2015low, iranzo2018probing,principi2018confining,lee2020image} since it will open up new opportunities to develop advanced optoelectronic devices operating at mid-infrared frequencies such as metasurfaces\cite{carrasco2015gate,huidobro2016graphene,biswas2018tunable}, photodetectors\cite{lundeberg2017thermoelectric, guo2018efficient}, and biosensors\cite{li2014graphene,rodrigo2015mid,hu2016far, hu2019gas}. Acoustic plasmons\cite{hwang2009plasmon,christensen2012graphene,pisarra2014acoustic,stauber2014plasmonics,alonso2017acoustic,iranzo2018probing,lee2019graphene,epstein2020far}, the hybridized bonding plasmon modes that are supported by spatially separated 2D materials, offer a practical route to push the light confinement toward its ultimate limit. In contrast to its counterpart with symmetric charge distributions (anti-bonding), the antisymmetric charge distributions of acoustic plasmons between two 2D layers help confine most of electromagnetic energies within the interlayer gap. As a result,  the plasmon confinement can be pushed to the extreme limit, beyond that of conventional 2D plasmons, as defined by the separation between the 2D layers. 
\par To date, acoustic graphene plasmons have been observed mostly in a graphene-on-a-mirror setup, where a single layer of graphene is placed near a metal film\cite{alonso2017acoustic, iranzo2018probing, lee2019graphene}. Hence, the graphene layer is paired with its image within the metal film due to an electromagnetic mirroring effect. Since the mirror image of the real graphene plasmon has an opposite charge distribution, the coupled graphene-metal system can only support the acoustic plasmons while the optical plasmons become the dark modes instead. In the double-layer system, on the other hand, optical plasmons are preferably excited due to the symmetry of the system while acoustic plasmons become difficult to excite. Therefore, experimental observation of acoustic plasmons in the double layer systems has been challenging.
\par In this work, we show how to selectively excite acoustic vs. optical plasmon modes in the double-layer graphene system. We develop a theory to calculate plasmon excitation by a single or multiple nano-emitters of deep sub-wavelength dimensions. Building upon the physical understanding facilitated by the theory, we design a far-field resonator that can selectively and efficiently excite the acoustic plasmons in the graphene double-layer system, which would allow us to reach the ultimate monoatomic-layer limit of the plasmon confinement inaccessible with the graphene-on-a-mirror system. 

\section{Mathematical formulation}

\begin{figure*}
\includegraphics[scale=0.9]{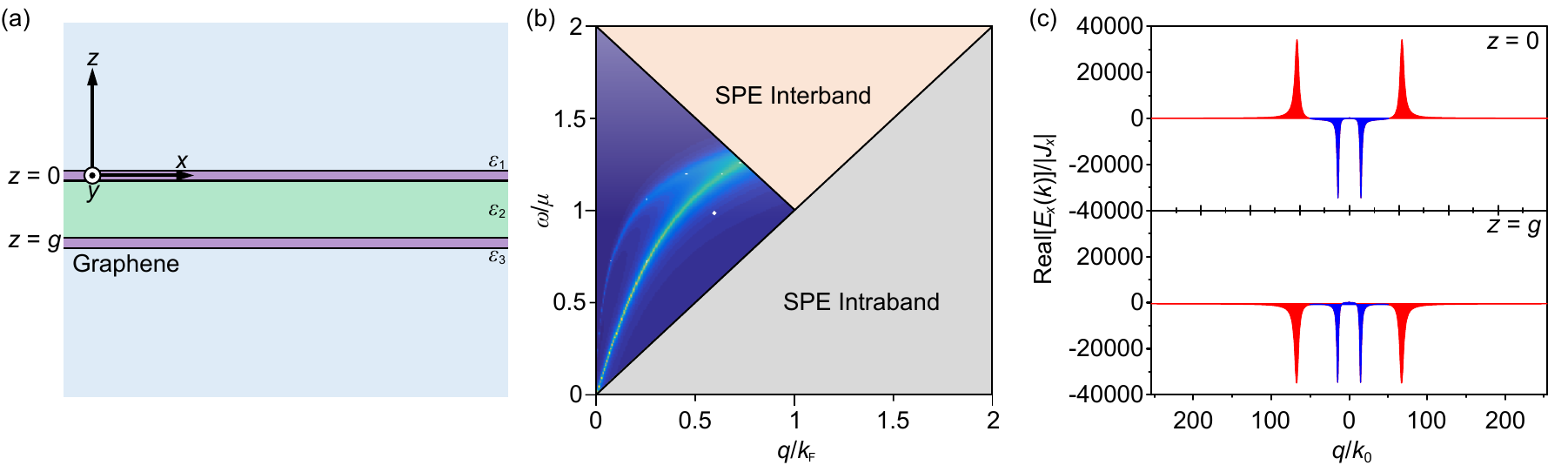}
\caption{\label{fig:wide}\textbf{Symmetry of graphene plasmons.} (a) Schematic illustration of a graphene double-layer system. A horizontally polarized Hertzian dipole is placed at a distance of h from the upper graphene layer located at $z=0$ (b) The plasmon dispersion of the double-layer system. 'SPE' stands for the single particle excitation. (c) The Fourier spectra of graphene plasmons excited with a single emitter placed 10 nm away from the upper graphene layer.}
\end{figure*}

Let us consider two infinite graphene sheets separated by a distance of $g$ as shown in Fig. 1(a). The optical conductivities $\sigma$ of the two electrically decoupled graphene sheets are calculated using the local approach \cite{nikitin2011fields} at a doping level $\mu$ of 0.4 eV and a damping rate $\eta$ of 10 meV. The plasmons are excited with a Hertzian dipole oscillating in $x$ direction with a current density of $J_{x}$ placed at a distance of $h$ from the top graphene sheet. Assuming harmonic time dependence $e^{i\omega t}$, the forced wave equation in a medium of permittivity $\epsilon$ for the $x$ component of electric fields $E_x$ is given as        
\begin{equation}
 \displaystyle\frac{\partial^2 E_x}{\partial y^2} -\gamma^2 E_x =\displaystyle\frac{iJ_{x}\gamma^2}{\omega \epsilon}\delta (y-h),  %
\end{equation}
where $\gamma=\sqrt{q^2-k^{2}_0}$ with $q$ and $k_{0}$ being the in-plane momentum of the plasmon and free-space wave, respectively. The solutions for Eq.(1) are given as 
\begin{equation} 
E_x=\begin{cases}
& E_1 e^{\gamma_{\rm I} z}-\xi e^{-\gamma_{\rm I}|z-h|}, \hspace{0.5cm}\text{$(z < 0)$} \\
& E_{\rm 2p}e^{-\gamma_{\rm II} z}+E_{\rm 2n}e^{\gamma_{\rm II} z}, \hspace{0.5cm}\text{$(0<z<g)$}\\
& E_{3}e^{-\gamma_{\rm III} (z-g)}, \hspace{0.5cm}\text{$(g<z)$}\\
\end{cases}
\end{equation}
where $\xi=\frac{iJ_x \gamma_{\rm 1}}{2\omega\epsilon_{\rm I}}$. By applying electromagnetic boundary conditions at the interfaces $z=0$ and $z=g$, the electric fields $\mathbf{E}$ in the gap region are related with the sources $\mathbf{S}$ through the system matrix $\mathbf{H}$,    
\begin{eqnarray}
&\mathbf{H}\mathbf{E}=\mathbf{S},\\
&\begin{bmatrix} A & B\\ 
C & D \end{bmatrix}\begin{bmatrix} E_{\rm 2n}\\E_{\rm 2p} \end{bmatrix} =   \begin{bmatrix} 2\alpha_1 \xi e^{-\gamma_{\rm I}|h|} \\ 0 
 \end{bmatrix}, 
\end{eqnarray}
where $A = (\Pi_1+1)\alpha_1+1$, $B=(\Pi_1+1)\alpha_1-1$, $C=[(\Pi_2+1)\alpha_2-1]e^{-\gamma_{\rm II}g}$, and $D=[(\Pi_2+1)\alpha_2+1]e^{\gamma_{\rm II}g}$ with $\Pi_1=\frac{i\sigma\gamma_{\rm I}}{\omega\epsilon_{\rm I}\epsilon_0}$, $\Pi_2=\frac{i\sigma\gamma_{\rm II}}{\omega\epsilon_{\rm I}\epsilon_0}$,$\alpha_1=\frac{\gamma_{\rm II}\epsilon_{\rm I}}{\gamma_{\rm I}\epsilon_{\rm II}}$, and $\alpha_2=\frac{\gamma_{\rm III}\epsilon_{\rm II}}{\gamma_{\rm II}\epsilon_{\rm III}}$. 
The zeros for the determinant of $\mathbf{H}$ give the plasmon dispersion of a double-layer system which is given as 
\begin{equation}
\displaystyle\frac{[(\Pi_1+1)\alpha_1+1][(\Pi_2+1)\alpha_2+1]}{[(\Pi_1+1)\alpha_1-1][(\Pi_2+1)\alpha_2-1]}=e^{-2\gamma_{\rm II}g},
\end{equation}
Fig. 1(b) shows the plasmon dispersion in a graphene double-layer with the gap size of 10 nm and $\varepsilon_{\rm 1}=\varepsilon_{\rm 2}=\varepsilon_{\rm 3}=1$. 

From $E_{\rm 2p}$ and $E_{\rm 2n}$, $E_{1}$ and $E_{3}$ are given as 
\begin{eqnarray}
E_{1} = E_{\rm 2p}+E_{\rm 2n}-\xi e^{-i\gamma_{\rm I}|h|},  \\ 
E_{3} = E_{\rm 2p}e^{-\gamma_{\rm II}g}+E_{\rm 2n}e^{\gamma_{\rm II}g}.
\end{eqnarray}

The symmetries of the plasmons excited from a nanoemitter, $\chi$, are defined by how much (the $E_x$ fields) at the surfaces of the two graphene layers are in phase. The in-phase and out-of-phase plasmon contributions can be analyzed using the Fourier spectrum of $E_x$. For example, the Fourier spectra of $E_x$ at the two graphene surfaces, i.e., $z=0$(upper surface) and $z=g$(lower surface), excited with a horizontally polarized emitter at $h=$10 nm, are shown in Fig. 1(c). The peaks with larger momenta correspond to acoustic plasmon contributions while smaller momenta peaks come from optical plasmon contributions. The red-colored portion of the plot shows out-of-phase components while the blue-colored portion indicates in-phase component, which also agrees with the symmetry of the acoustic and optical plasmons. Here, we define the measure of acoustic plasmon content $\chi$ from the Fourier spectra of graphene plasmons excited with a nanoemitter as follows: 

\begin{equation}
 \chi_{\upsilon} := \frac{\int_{\frac{\pi}{2}<\phi(q)<\frac{3\pi}{2}}|E_x|dq-\int_{0<\phi(q)<\frac{\pi}{2}, \frac{3\pi}{2}<\phi(q)<2\pi}|E_x|dq}{\int_{\frac{\pi}{2}<\phi(q)<\frac{3\pi}{2}}|E_x|dq+\int_{0<\phi(q)<\frac{\pi}{2}, \frac{3\pi}{2}<\phi(q)<2\pi}|E_x|dq},
\end{equation}
where $\upsilon=1,2$ denote the case of $z=0$ and $z=g$, respectively and $\phi(q)$ is the relative phase between $E_x(z=0)$ and $E_x(z=g)$. The $\chi$ of the system is given as an average of the two $\chi$s calculated for the two graphene layer respectively, i.e.,  $\chi=\left(\frac{\chi_1+\chi_2}{2}\right)$. Note that $\chi\rightarrow1$ when acoustic plasmons dominate while $\chi\rightarrow -1$ when optical plasmons dominate.

\section{Single nanoemitter excitation}
The symmetry of plasmons excited by a single nanoemitter is dependent on the polarization of the nanoemitter and its relative position to the double layer. When a nanoemitter is placed far away from the system $(h \ll 0)$ or $(g \ll h)$, the optical plasmons are preferably excited irrespective of the polarization of the emitter since the fields radiated by the emitter reach the two graphene layers with similar magnitudes and phases (Figs. 2(a) and 2(b)). The asymmetric coupling has more to do with different near-field coupling to the two layers. However, even when the emitter is infinitesimally close to the graphene, $\chi$ is around 0.5, suggesting still significant contributions from the optical plasmons. On the other hand, $\chi$ can be effectively tuned between $\pm 1$ when the emitter is placed between the two graphene layers, i.e., $h=0$. In contrast to the case of horizontal polarization, where the radiated fields at the two graphene layers are symmetric, the use of a vertically polarized dipole mandates the radiated fields at the two graphene layers to be asymmetric, allowing to achieve $\chi$ of unity. The analytical results can be verified with numerical simulations of the nanoemitter-launched fields  carried out for the cases where the emitter with horizontal (Fig. 2(c)) and vertical (Fig. 2(d)) polarization is placed in the middle of the two graphene layers. Appendix A provides additional insights on the origin of the change in the sign of $\chi$ as a function of $h$, based on an analysis of the Purcell factor.        

\begin{figure}
\includegraphics[scale=0.97]{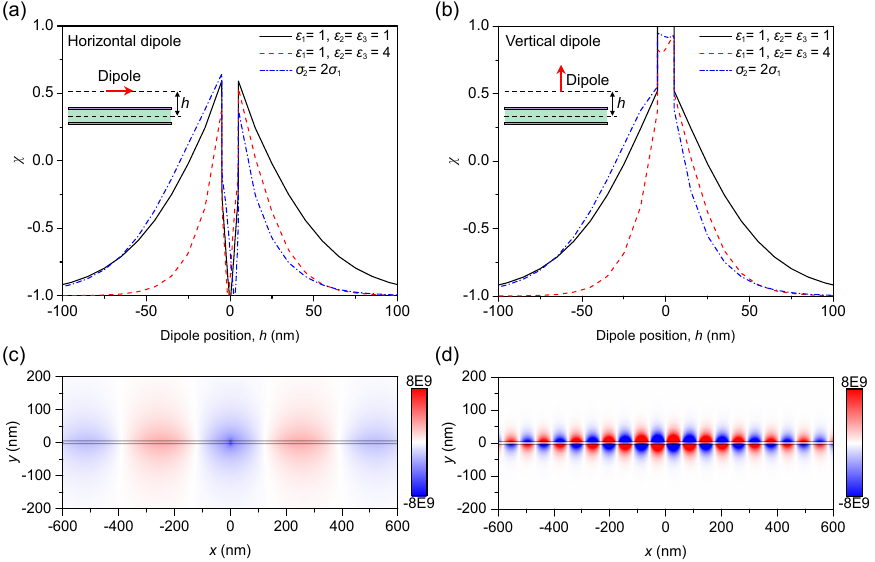}
\caption{\textbf{Plasmon excitation with a single emitter.} The symmetry of the graphene plasmons as a function of the distance of (a) a horizontally and (b) vertically oriented dipole from the top graphene layer. Spatial distributions of electric fields in the $x$ direction when $h=0$ excited with a (c) horizontally and (d) vertically oriented dipole.}  
\end{figure}

\section{Multiple nanoemitter excitation}
Although a single nanoemitter with vertical polarization placed between two graphene layers can exclusively excite acoustic plasmons, the implementation of such configuration is impractical due to the difficulty in embedding a nanoemitter between the two layers as well as the precise orientation of the nanoemitter. Here we demonstrate that multiple nanoemitters can excite acoustic plasmons with high efficiencies irrespective of the polarization of the emitter, making it more attractive for practical implementation. The case of two emitters gives an insight on how multiple emitters can help control the value of $\chi$. Let us consider the case of both emitters placed a few nanometers above the double-layer system. As the spacing between the two emitters $s$ increases, the relative phases between plasmons excited from the two dipoles $e^{iqs}$ change leading to the oscillatory behavior of $\chi(s)$ as shown in Fig. 3(a). $\chi$ is maximized when the constructive interference condition $qs = 2m\pi$ is met with $m$ being an integer. The plasmon damping limits the maximum achievable values for $\chi$. 
\par From the two emitter case, it is obvious that a periodic arrangement of multiple emitters can revoke the constructive interference condition periodically and increase the acoustic contributions to plasmons excited from the emitters (Fig. 3(b)). As the number of emitters $N$ arranged with a periodicity of $s$ increases, $\chi$ asymptotically approaches 0.94, which is limited by the plasmon damping. $\chi$'s for the horizontal and vertical case are identical, which shows the robustness of the multiple-emitter-based approach. The near-field distributions from numerical simulations show that the acoustic contributions prevail for $N$ = 21 (Figs. 3(c) and 3(d)).

\begin{figure}
\includegraphics[scale=0.97]{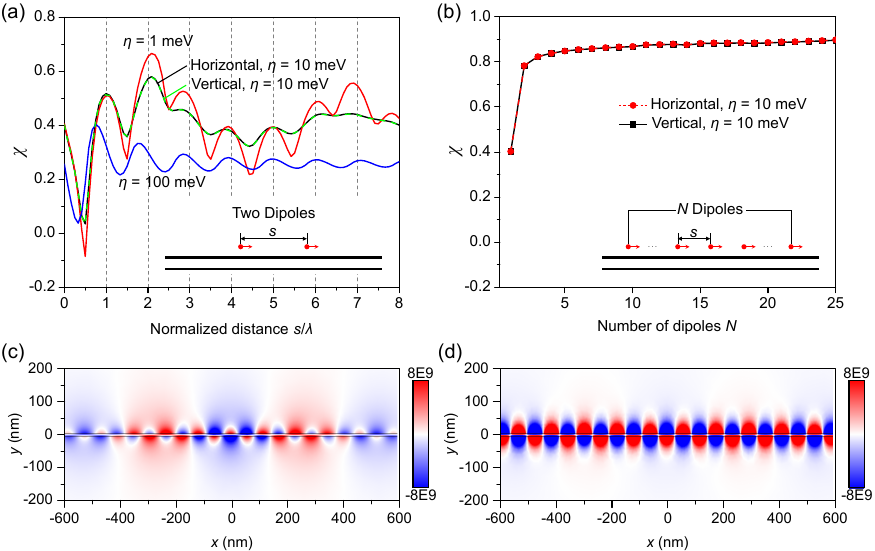}
\caption{\textbf{Plasmon excitation with multiple emitters} (a) The symmetry of the graphene plasmons as a function of the spacing between two dipoles. (b) The symmetry as a function of the number of dipoles separated by 120 nm. Spatial distributions of electric fields in the $x$ direction when (c) $N=2$ and (d) $N=21$. }
\end{figure}

\section{Efficient far-field excitation of acoustic graphene plasmons }
Motivated by the multiple-emitter-based approach, we suggest a resonator design that can excite acoustic plasmons with high efficiency and selectivity. The narrow slits in the graphene layer can play as near-field emitters by scattering far-field radiation. By introducing a periodic array of such narrow slits in a graphene layer, multiple emitters discussed in the previous section can be effectively implemented with far-field excitation. In the resonator design consisting of a pair of identical graphene ribbon arrays shown in the upper panel in Fig. 4(a), the far-field radiation reaches the narrow slits in the upper and lower graphene with similar magnitudes and phases, which is equivalent to the case of having each of two identical emitters on the upper and lower graphene layer. In this case, the optical plasmons are preferably excited while the acoustic plasmons remain a dark mode. On the other hand, the asymmetric design presented in the lower panel consists of a graphene ribbon layer separated by narrow slits and a continous graphene sheet to realize the multiple-emitter design previously discussed.  
\par The far-field spectra of the two resonator designs are drastically different as shown in Fig. 4(b). The difference can be attributed to the distinct modal natures for the two designs. The near-field distribution ($E_x$) for the resonance from the double layer of graphene ribbon arrays shows that it  originates from the optical plasmons (Fig. 4(c)). On the other hand, the resonance at the lower frequency for the asymmetric design shows that the resonance is dominated by the acoustic plasmons (Fig. 4(d)), demonstrating that such design indeed efficiently couple far-field radiation to the acoustic plasmons.

\begin{figure}[h]
\includegraphics[scale=0.97]{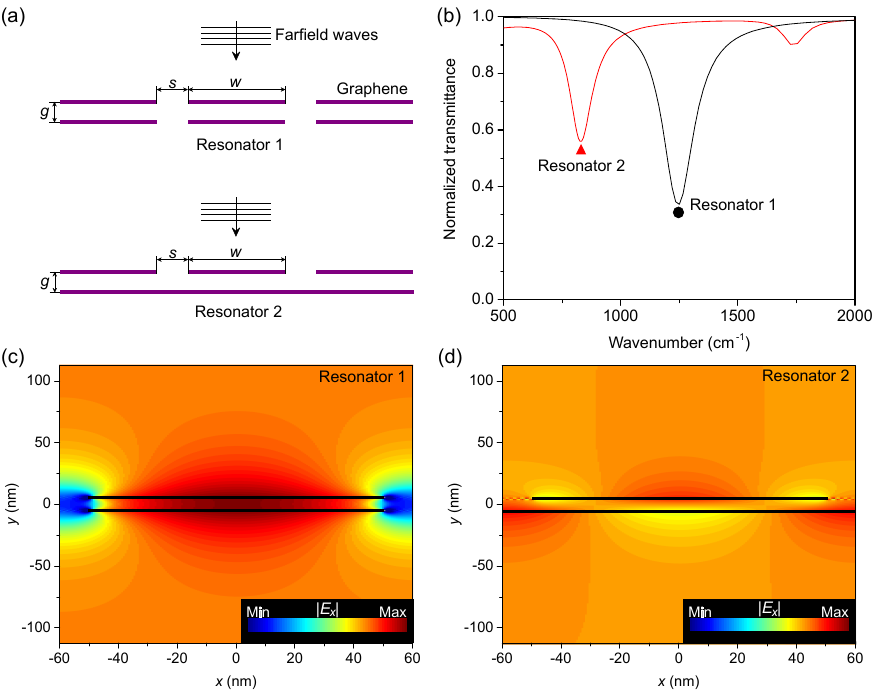}
\caption{ \textbf{Far-field excitation of acoustic plasmons} (a) (Upper) Resonator design based on the double graphene ribbon arrays. (Lower) New resonator design based on a continuous graphene sheet with a graphene ribbon array. (b) Numerical results for far-field spectra obtained from the two designs. The red curve shows the case of the new design based on a continuous graphene layer. The spatial distributions of electric fields in the $x$ direction for (c) double ribbon arrays  and (d) asymmetric double-layer design.}
\end{figure}

\section{Ultimate plasmon confinement}
The excitation of the acoustic plasmons in the double-layer graphene system allows for probing the fundamental limit of the graphene plasmon. The lateral confinement of the graphene plasmon as measured by its effective index, $q/k_0$, increases with decreasing $g$. Therefore, the ultimate limit can be achieved with a mono-atomic film such as a monolayer of hexagonal boron nitride (hBN)\cite{dean2010boron} as shown in Fig. 5(a). The same strategy has been experimentally demonstrated with the conventional graphene-metal coupled architecture with a monolayer of hBN in-between as shown in Fig. 5(b). Here, the inter-layer separation is effectively double-atomic-layer thickness due to the electromagnetic mirroring effect. On the other hand, our graphene double-layer system can access the fundamental limit of the lateral plasmon confinement, i.e., monoatomic-layer thickness (Fig. 5(c)). In practice, the nonlocal effects arising from the nonlocal conductivities, which can be calculated in the random phase approximation\cite{hwang2007dielectric}, limit the achievable plasmon confinement\cite{lundeberg2017tuning, dias2018probing, iranzo2018probing}. 

\begin{figure}
\includegraphics[scale=0.97]{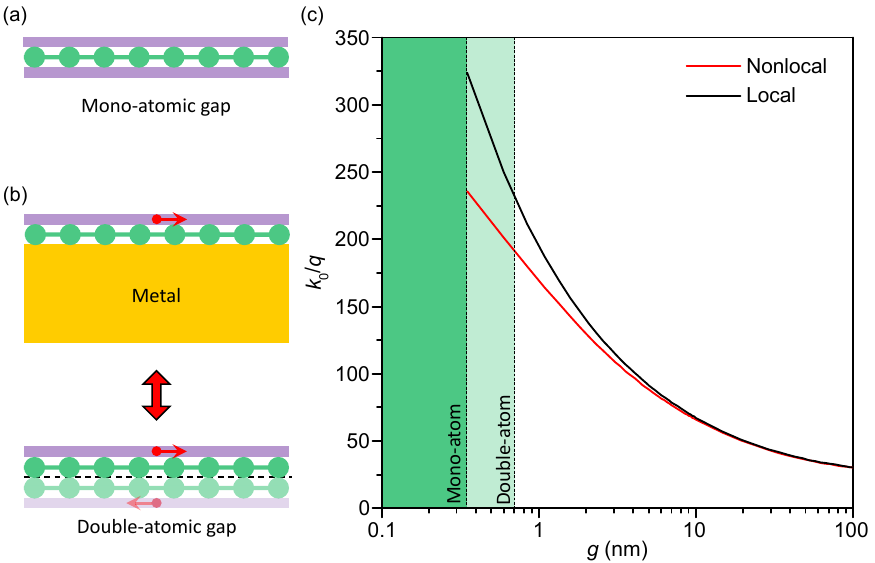}
\caption{\textbf{Ultimate monoatomic plasmon confinement limit.} (a) Schematic illustration showing two graphene layers separated by a monoatomic spacer. (b) Schematic illustration of the graphene-on-a-mirror system with a monoatomic spacer in between. This configuration is effectively equivalent to that of double-layer graphene separated by a double-atomic-layer spacer. (c) The lateral plasmon confinement as a function of $g$ at $\lambda=$ 8 $\mu$m.}
\end{figure}

\section{Conclusions}
We have demonstrated that the symmetry of the graphene plasmons can be switched with a judicious geometrical arrangement of a single or multiple nano-emitters. Our analytical theory allows us to optimize the excitation scheme to efficiently and selectively launch acoustic graphene plasmons from far-field radiation. Also, the analytical results inspired us to design acoustic graphene plasmon resonators consisting of a continuous graphene layer and a graphene ribbon array. Our practical resonator design will enable experimental observation of the ultimate level of plasmon confinement defined by a monoatomic layer, which is inaccesible with a conventional graphene-metal coupled architecture. Also, the efficient and robust excitation of the acoustic plasmons will benefit fundamental studies such as nonlocality and nonlinearity\cite{jiang2019ultrafast} as well as a variety of applications including active metasurfaces\cite{han2020complete}, biosensors\cite{chen2017acoustic}, and photodetectors\cite{koppens2014photodetectors}.

\begin{acknowledgments}
This research was supported by grants from the U.S. National Science Foundation (NSF) MRSEC Seed (to I.-H.L., T.L., and S.-H.O.), NSF ECCS Award No. 1809723 (to I.-H.L., S.-H.O., T.L.), and the Samsung Global Research Outreach (GRO) Program (to S.-H.O.). L.M.-M. acknowledges Spain’s MINECO under Grant No. MAT2017-88358-C3, funding from the European Union Seventh Framework Programme under grant agreement No. 881603 Graphene Flagship for Core3, and Aragon Government through project Q-MAD. S.-H.O. further acknowledges support from the Sanford P. Bordeau Chair in Electrical Engineering at the University of Minnesota.  
\end{acknowledgments}

\appendix

\section{Purcell factor}

How different plasmons are excited can be characterised by the Purcell factor. For a one-dimensional line of vertical dipoles, placed at a position $h$, the Purcell factor is related to the reflection coefficient at a give in-plane wavevector $r(q)$ by

\begin{equation}
 P = 1-4\Re\left(\int_{-\infty}^{\infty} \frac{r(q)}{2k_z}e^{2ik_{z}z'}\frac{k^2_x}{{k_0}^2} dk_x\right),
\end{equation}

where $k_z = \sqrt{k_0^2 - q^2}$, with $k_0 = \omega/c$ (note that we have written down the exact expression, not the one within the quasi-static approximation).  
The poles of $r_x$ provide the dispersion relation of bound modes (in this case, the acoustic and optical plasmons). We denote the contribution of these plasmons to the Purcell factor  as $P_1$ (the least confined mode) and $P_2$ 
(the more confined one). They can be computed by fitting the corresponding reflection resonances to Lorentzians, which can be analytically computed.
Correspondingly, the acoustic plasmon content $\chi$ can be expressed as $\chi_P=(P_2-P_1)/(P_2+P_1)$.

\begin{figure}
\includegraphics[scale=1.05]{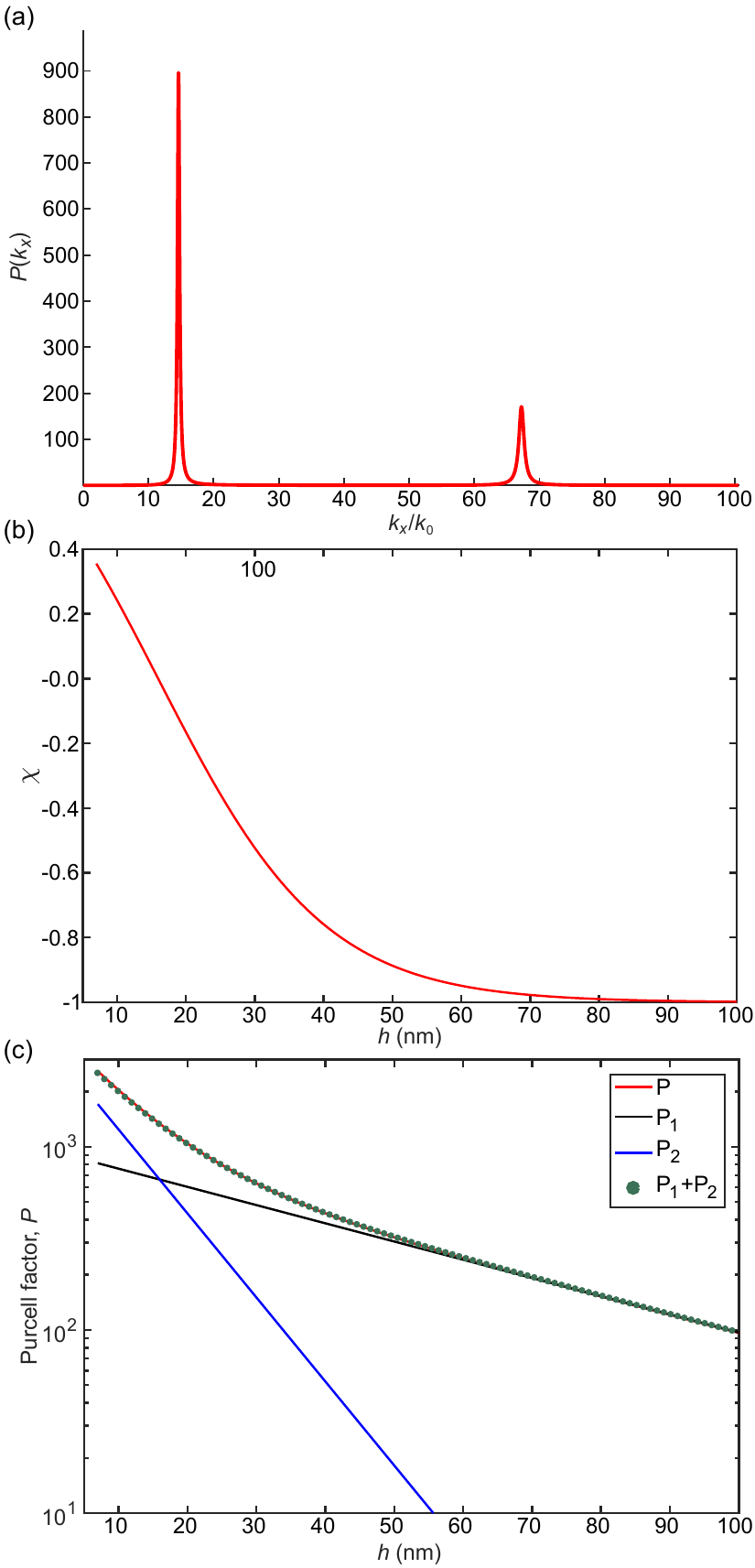}
\caption{\textbf{Purcell factor.} (a) Then integrand of Eq. (A1) as a function of in-plane wavevector for the case $h=10$ nm. 
(b) The dependence of $\chi$ with distance $h$. (c) The total Purcell factor, together with $P_1$ and $P_2$, as function of the distance 
between the line of dipoles and one of the graphene layers.}  
\end{figure}

Figure 6 illustrates the Purcell factor and the different contributions, for the representative values: chemical potential $\mu = 0.4 $ eV,  temperature $T = 300$ K and wavelength $\lambda= 8 \mu$m.

Fig. 6(a) shows the integrand of Eq. (A1) as a function of in-plane wavevector, for the case $h=10$ nm, showing clearly that virtually all the dipole radiation goes into two well-defined plasmonic modes, with wavevectors $k_{p1} \approx 14.6 k_0$ and $k_{p1} \approx 67.2 k_0$.
Fig. 6(b) shows the dependence of $\chi_P$ with distance $h$.
Fig. 6(c) shows the total Purcell factor, together with $P_1$ and $P_2$, as function of the distance 
between the line of dipoles and one of the graphene layers.
The figure shows that at small distances the field radiated by the dipole couples preferentially to the most confined mode (the acoustic plasmon), but the as $h$ grows the coupling to the acoustic plasmon decreases faster than the one to the optical plasmon, which dominate the Purcell factor for distances $ h\approx 100$ nm. This evolution explains the dependence of $\chi_P$ with distance $h$ rendered in Fig. 6(b).

\bibliographystyle{apsrev4-2}
\bibliography{main}

\end{document}